# Large transport critical currents and magneto-optical imaging of textured $Sr_{1-x}K_xFe_2As_2$ superconducting tapes

Yanwei Ma*, Chao Yao, Xianping Zhang, He Lin, Dongliang Wang

Key Laboratory of Applied Superconductivity, Institute of Electrical Engineering,

Chinese Academy of Sciences, Beijing 100190, China

A. Matsumoto, H. Kumakura

National Institute for Materials Science, Tsukuba, Ibaraki 305-0047, Japan

Y. Tsuchiya, Y. Sun, T. Tamegai

Department of Applied Physics, the University of Tokyo, Hongo, Tokyo 113-8656, Japan

**Abstract:**

We report the temperature dependence of the transport critical current density ($J_c$) in textured $Sr_{1-x}K_xFe_2As_2$/Fe (Sr122) tapes fabricated by an *ex situ* powder-in-tube process. Critical currents were measured in magnetic fields up to 0-15 T and/or the temperature range 4.2-30 K by using a dc four-probe method. It was found that textured Sr122 tapes heat-treated at low temperatures showed higher transport $J_c$ performance due to much improved intergrain connections. At temperatures of 20 K, easily obtained using a cryocooler, $J_c$ reached ~ $10^4$ A/cm$^2$ in self field, which is the highest transport value of ferropnictide wires and tapes reported so far. Magneto-optical imaging observations further revealed significant and well distributed global $J_c$ at 20 K in our tapes. These results demonstrate that 122 type superconducting tapes are promising for high-field applications at around 20 K.

---

* E-mail: ywma@mail.iee.ac.cn

**Introduction**

The K doped Ba- or SrFe$_2$As$_2$ (122 type) superconductors have the advantages of extremely high upper critical field ($H_{c2}$) of about 40-50 T at 20 K [1, 2] and very low anisotropy $1 < \gamma < 2$. Especially, large critical current densities ($J_c$) of $10^5$-$10^6$ A/cm$^2$ with almost independent of the field have been reported in 122 films grown on metal substrates [3, 4]. These results strongly suggest that the conductors based on 122 compounds are more promising for high field magnets operating at 20 K [5], where the niobium-based superconductors cannot play a role owing to their lower $T_c$s, and MgB$_2$ wire is also difficult to be employed because of a very rapid drop of $J_c$ with the applied field.

Soon after the discovery of pnictides [6], an attempt has been made to realize a wire conductor that would lead to potential applications [7, 8]. The method commonly used to fabricate iron-based superconducting tapes is the powder-in-tube (PIT) technique [5]. There are two main variants of PIT pnictide fabrication, *in situ* [7-9] and *ex situ* [10-12]. Most groups have focused on the development of *ex situ* wires because of perceived advantages in both fewer impurity phases and a high density of the superconducting core.

To date, enormous efforts have been directed towards improving the in-field $J_c$ of 122 pnictide wires and tapes through the development and application of various methods for fabrication of technically usable materials, such as chemical doping [10, 11, 14], hot isostatic pressing (HIP) [13] and texturing technique [14]. At 4.2 K, 0 T, high $J_c$ values of $10^4$ to $10^5$ A/cm$^2$ have been reported for 122 wires and tapes. More recently, we have demonstrated that transport $J_c$ values above $10^4$ A/cm$^2$ in a high field of 10 T were achieved in Sr122/Fe tapes through a texturing process plus Sn addition, which are both effective ways to improve grain connectivity [15]. It is noted that quantification of transport measurements reported so far is almost performed at 4.2 K for convenience.

On the other hand, 122 wires possess a strong potential for high field application at ~ 20 K due to their high $H_{c2}$ at this temperature. In order to examine the feasibility of 122 wires at high temperatures for a real application, it is important to have a complete transport $J_c$ data performed as a function of temperature and field. In this study, we report our findings regarding the temperature dependence of the transport critical current properties and the irreversible field of textured Sr122 superconducting tapes. We also used a magneto-optical imaging technique to investigate the properties of grain boundaries in the Sr122 tapes.

**Experimental**

Textured $Sr_{0.6}K_{0.4}Fe_2As_2$/Fe (Sr122) composite tapes were prepared by the *ex situ* powder-in-tube (PIT) process. The details of fabrication process have been reported in details in [16]. In order to compensate for loss of elements during the milling and sintering procedures, the starting mixture contains 10-20% excess K [17]. The precursors with the addition of 5 wt% Sn were ground. The final powders were filled and sealed into an iron tube, which was subsequently swaged and drawn down to a wire of 1.9 mm in diameter. The as-drawn wires were then cold rolled into tapes (0.6 mm in thickness) with a reduction rate of 10~20%. Short samples (4 cm each), cut from the tapes, were finally sintered following two different processes: i) Samples were heated at 1100$^o$C for 1~5 minutes (high temperature annealing process: HT tape); ii) Samples were sintered at 800-950$^o$C for 5~30 minutes and then decreased to 600$^o$C for 5 h (low temperature annealing process: LT tape).

Transport measurements were performed, at the High Magnetic Field Station, National Institute for Materials Science (NIMS), by the standard four-probe method using 18 T superconducting magnet. Magnetic fields were applied parallel to the tape surface. The sample temperature was controlled with helium gas cooling. Critical current ($I_c$) values at 4.2 K were obtained from the measurements in liquid helium. At high temperatures, the measurements were performed with stabilizing sample temperature within the error of 0.1 K by using helium gas cooling. $I_c$ was determined from the current-voltage curve with 1 µV/cm electric field criterion. $J_c$ was estimated by dividing the $I_c$ by the cross sectional area of the core of the tape. The DC magnetization of samples was measured using a Quantum Design physical property measurement system (PPMS) at 5 K and 20 K, respectively. Magneto-optical (MO) imaging was done at the University of Tokyo. A Bi-substituted iron-garnet indicator film was placed directly onto the sample surface, and the whole assembly was attached to the cold finger of a He-flow cryostat (Microstat-HR, Oxford Instruments). MO images were obtained by using a cooled CCD camera with 12-bit resolution (ORCA-ER, Hamamatsu).

**Results and discussion**

Data shown in figure 1 is based on transport measurements performed on a HT tape that had experienced a very short high temperature at 1100$^o$C. Transport $J_c$ measurements on this sample were made over a range of temperatures (10-30K). Several conclusions can be drawn from figure 1. First, $J_c$ decreased with increasing temperature (with increasing field). Second, the field dependence of the $J_c$ became weaker with decreasing temperature. Even this, the measured 122 tapes exhibited more independent of field behavior at high temperatures, e.g., at 20 K, compared to MgB$_2$ tapes, where the $J_c$ usually dropped rapidly with the applied field. Third, At 10 and 15 K, $J_c$ values were over $10^4$ A/cm$^2$ at 0 T, for example, at 10 K, $J_c$ possessed ~ $2\times10^4$ A/cm$^2$ in self field, and still remained ~3000 A/cm$^2$ until 15 T. While $J_c$ at 20 K was 5700 A/cm$^2$ in self field and 230 A/cm$^2$ at 10 T, respectively.

Figure 2 shows the magnetic-field dependence of the transport $J_c$ at various measuring temperatures for the LT processed tapes, which were heat treateded at low temperatures ranging from 850 to 950$^o$C but for longer time. The superiority of the LT samples is immediately clear, namely, higher $J_c$ values at different measuring temperatures were achieved in the LT tape in terms of HT samples. At 20 K, the transport $J_c$ reached ~$10^4$ A/cm$^2$ at 0 T and ~650 A/cm$^2$ at 10 T, respectively. Even measuring at 25 K, we observed $J_c$ values of over 2100 A/cm$^2$ at 0 T and 160 A/cm$^2$ at 5 T, respectively. The large $J_c$ values in the high field range can be due to the high $H_{c2}$ values, as discussed below. It is noted that the $J_c$-$H$-$T$ properties of our LT Sr122 tapes are larger than the *ex situ* Ba$_{0.6}$K$_{0.4}$Fe$_2$As$_2$ wires made by Ag addition [18]. These promising results at high temperatures suggest that the textured Sr122 superconducting tapes will have a strong potential in high field magnets operating at 20 K.

The same data, displayed as the temperature dependence of $J_c$ at zero field for the LT and HT tapes, are presented in the inset of figure 2. The $J_c$ values of both tapes increased rapidly with decreasing temperature down to 15 K and tend to saturate with further the decrease of temperature. Evidently, the $J_c$ of the LT tapes rose more rapidly with lowing temperature than that of the HT ones. A possible explanation for these results is ascribed to improved grain connections of LT tapes because of dense structure and reducing the inhomogeneities during the low temperature process [15].

Figure 3 shows $J_c$ for the LT tape as a function of magnetic field at 5 and 20 K, as calculated from magnetization loops M(*H*) using the critical state Bean model [19]. Clearly, the magnetic $J_c$s, measured on short sections, were higher than the transport results, indicating that the intragrain $J_c$ component still makes a considerable contribution to the magnetization hysteresis, especially at low temperature region. Another likely reason for the discrepancy is the possible presence of inhomogeneities along the length of the tape, which would result in suppressed transport $J_c$.

Magneto optical (MO) imaging was used to examine grain boundary features and then established the correspondence between these features and the observed transport $J_c$ behavior. Figures 4(a) shows the optical image of the polished core surface of the LT tape, note that the polished core surface is parallel to the tape surface. Figures 4(b)-(e) reveal the penetration of magnetic flux after zero-field cooling (ZFC) the sample to 20 K. As the external field increases, flux starts to penetrate at ~5 mT and only a partial flux penetration can be visible due to strong shielding currents caused by applying a magnetic field even at ~50 mT, indicative of the bulk circulating currents in our tape.

After the removal of the external field of 80 mT at different temperatures, the remanent trapped fields are shown in figures 4(f)-(h). The bright region corresponds to the trapped flux in the sample. When the temperature increases, the bright area progressively approaches the center of the sample, implying that the long range current decreases as the temperature is increased toward $T_c$. Most interestingly, the trapped fields at various temperatures were mostly homogenous after applying a high external field of 80 mT. Furthermore, figure 4(h) clearly showed a remarkable 'roof pattern' at 20 K and testifying the almost uniform bulk-scale current flow in our sample. These images seem comparable to the MO images of HIP processed Ba122 wires [13], but much better than those of other pnictide polycrystals and wires showing primarily granular currents indicating little or no bulk current flow [12, 20, 21]. The estimation from the magnetic profile in figure 4(h) gives us the value of the global current $J_c$ of the order of 2-3×10$^4$ A/cm$^2$ at 20 K, which is the same order of magnitude of transport $J_c$ by the four-probe method. Obviously, the correspondence between MO and transport results is evident.

In order to evaluate the $H_{irr}$ values at elevated temperatures, we performed the resistance versus temperature measurements on the LT processed Sr122 tapes in different magnetic fields by the four-probe resistive method, as shown in the inset of figure 5. With increasing magnetic field, both the onset and zero resistive points of the superconducting transition curve shifted towards low temperatures, similar to the case of MgB$_2$ tapes [22], however, both transition temperatures were not sensitive to magnetic field, an indicative of less weak-linked behavior in our textured Sr122 tapes. Figure 5 shows the change of transition temperature with the critical field ($H_{c2}$ and $H_{irr}$) for LT processed Sr122 tapes. The $H_{c2}$ and $H_{irr}$ are defined by using two criteria, the 90% and 10% values of the normal-state resistance, respectively.

From figure 5, we can immediately notice that both $H_{c2}$ and $H_{irr}$ curves of Sr122 tapes are much steeper than those of MgB$_2$ tape samples [22]. For instance, $H_{irr}$ of MgB$_2$ tapes reaches 9 T at only ~21 K, while $H_{irr}$ of 9T can be achieved for Sr122 tapes at higher temperature up to ~31 K. The larger irreversibility fields at high temperatures do support our transport measurements (see figures 1 and 2) that Sr122 tapes exhibited weaker field-dependent $J_c$, which is a big advantage in view of the fabrication of magnets operating at 20 K. In addition, the slope of $H_{c2}$ at $T_c$ is 4.55 $T$/K. The value of $H_{c2}$ at $T$=0 K estimated using the Werthamer–Helfand–Hohenberg formula [23], 0.693×(d$H_{c2}$/d$T$) ×$T_c$, is ~110 T, very comparable to that of the single crystal [24].

Compared to our earlier studies of *ex-situ* PIT Sr122 tapes [10], the texture processing of the present samples seemed to increase the number of local links and strong links, leading to better connectivity and thus more intergranular current loops.

However, the low temperature sintering process (LT tape) resulted in highly dense material with less impurity phases, which contributed to good connection. As a result, the LT tapes had higher transport $J_c$ in the whole temperature and field ranges, reflecting the better performance in $H_{irr}$, as displayed in figure 5. At 20 K, the transport $J_c$ reached ~$10^4$ A/cm$^2$ at zero field and ~650 A/cm$^2$ at 10 T, respectively, demonstrating the potential for high field magnet applications. This is in agreement with the recent results of [18] where the transport $J_c$ at 20 K is weakly affected by the applied field. Additionally, MO imaging further demonstrates there is well distributed global $J_c$ in the present samples. We expect that these data can be useful for choosing applications and operational regimes for 122 pnictide tapes.

On the other hand, compared with the magnetic and transport $J_c$ at 5 and 20 K, the magnetic $J_c$ (and also the estimated $J_c$ from the MO image) was larger than the transport $J_c$. This is in accordance with observations by several groups who noted the hysteresis between $J_c$-$H$ curves in a cycling applied magnetic field at different temperatures [9, 18, 25, 26], in other words, the $J_c$ value differs when measured in increasing and decreasing fields. Those data clearly suggest that the current path is still limited in the tape core and grain boundaries still act as a block to some extent. Therefore, it is still very necessary to control grain misorientations by developing high texture to obtain high intergranular $J_c$ in pnictide wires and tapes, similar to Bi-2223 cuprate. Based on our present results, further improvement in $J_c$ is expected upon either utilization of high purity 122 phases or optimization of the texturing process.

**Conclusions**

High temperature performance was investigated for textured Sr-122 tapes/Fe by *the ex-situ* PIT technique. $I_c$-$T$-$H$ relations were also examined. The combined magnetic, magneto-optical and transport analyses showed that textured PIT processing enhanced the intergranular current, hence the large $J_c$. At 20 K, $J_c$ value of ~ $10^4$/cm$^2$ was achieved for LT Sr122 tapes. Nevertheless, the $J_c$ at 20 K obtained in this study is the highest value measured among wires and tapes. If we succeed in further enhancing the $J_c$, these results demonstrate that 122 type superconducting tapes are promising for high-field applications at around 20 K.

**Acknowledgements**

The Authors thank S. J. Ye at NIMS for his help during the experiments. This work is partially supported by the National '973' Program (Grant No. 2011CBA00105) and National Natural Science Foundation of China (Grant No. 51025726 and 51172230), and China-Japan Bilateral Joint Research Project by the JSPS and NSFC.

**Captions**

Figure 1 Transport $J_c$ versus applied magnetic field at temperatures between 10 and 25 K for the HT Sr122 tapes

Figure 2 Magnetic field dependence of the transport $J_c$ at various measuring temperatures for the LT processed tapes. The inset shows the temperature dependence of $J_c$ for the LT and HT tapes.

Figure 3 Magnetic field dependence of $J_c$ for the LT processed samples at 5 and 20 K, as measured by magnetization.

Figure 4 (a) Optical image of the polished core surface of the LT tape. MO images of magnetic flux penetration into the sample at (b) 5 mT, (c) 20 mT, (d) 30 mT, and (e) 50 mT after zero-field cooling down to 20 K. Different MO images in the remanent state of the sample at (f) 5 K, (g) 10 K, and (h) 20 K after cycling the field up to 80 mT.

Figure 5 Upper critical field lines $H_{c2}$ and $H_{irr}$ as a function of the temperature for the LT Sr122 and $MgB_2$ tapes. The $H_{c2}$ and $H_{irr}$ values were defined as the 90% and 10% points of the resistive transition, respectively. The inset shows resistivity at different fields.

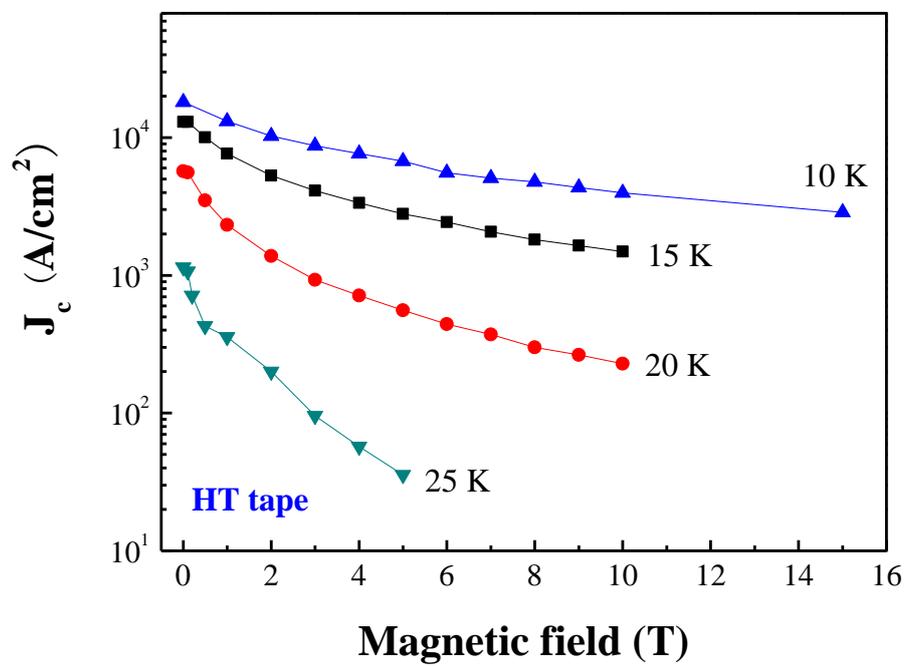

Figure 1   Ma et al.

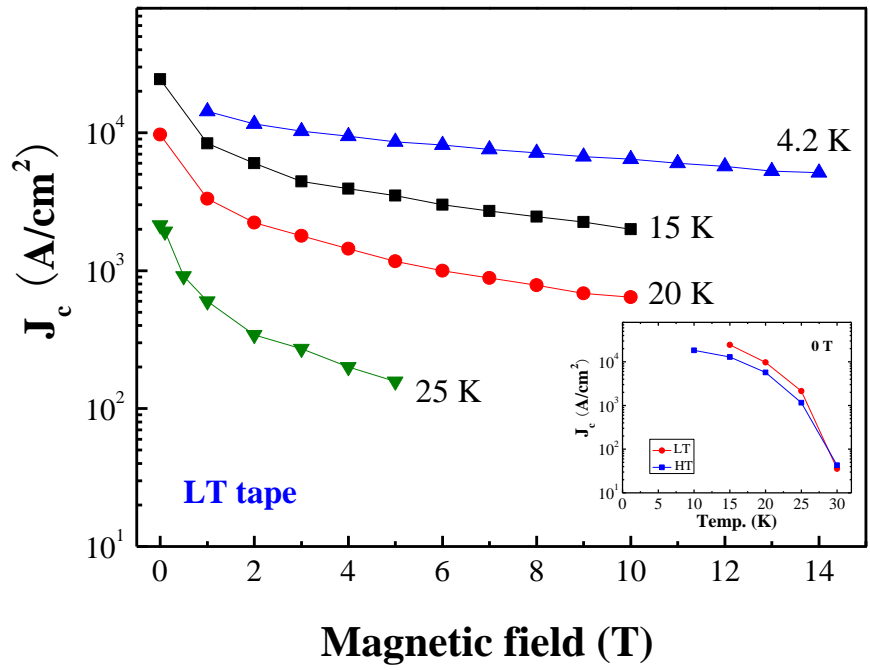

Figure 2    Ma et al.

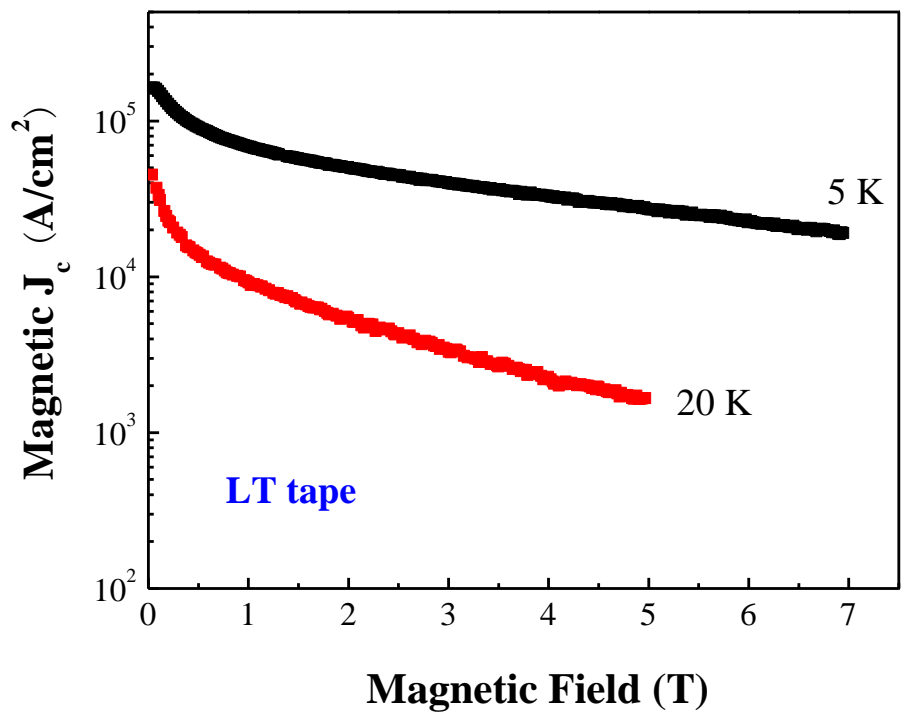

Figure 3    Ma et al.

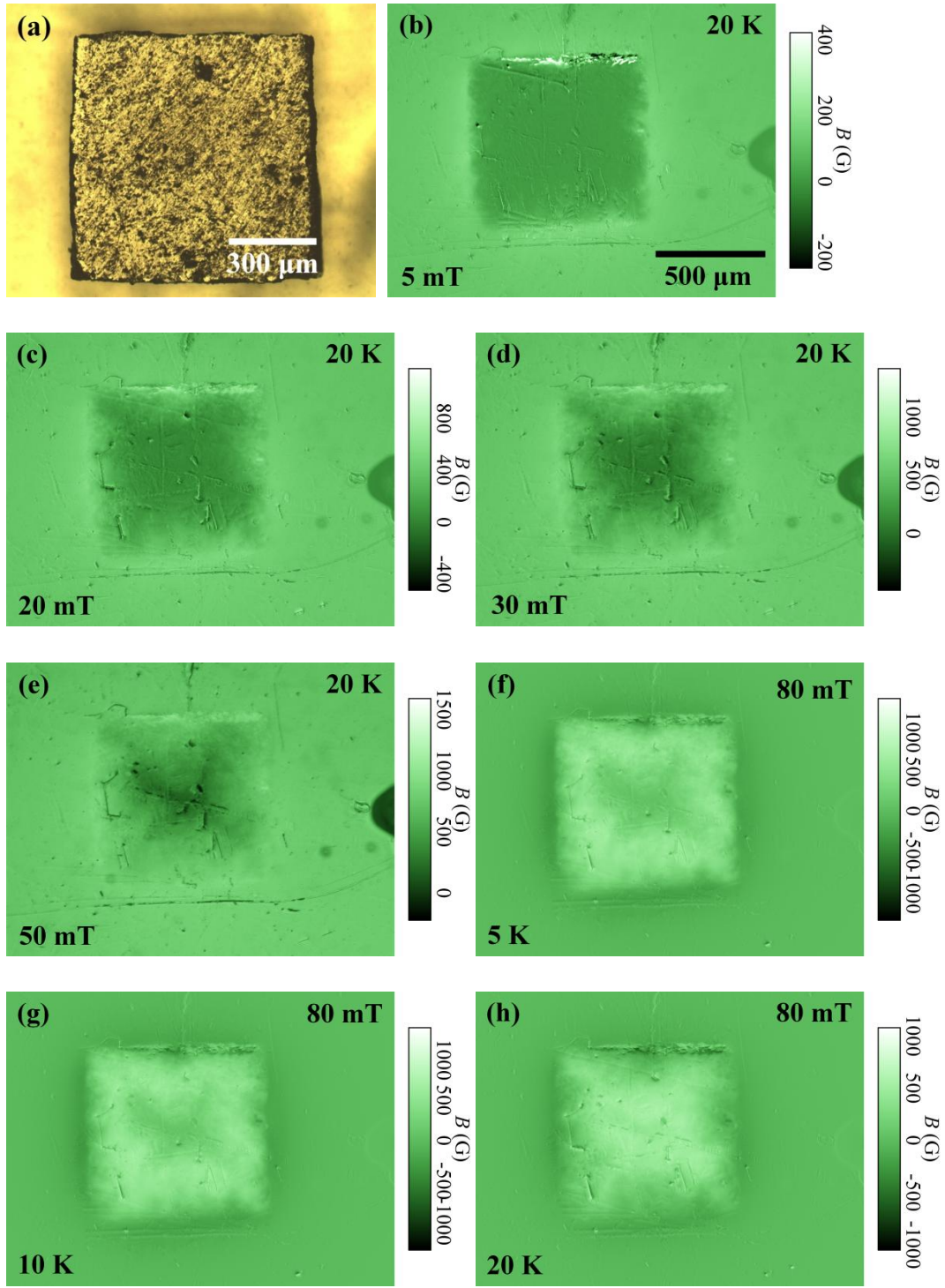

Figure 4 Ma et al.

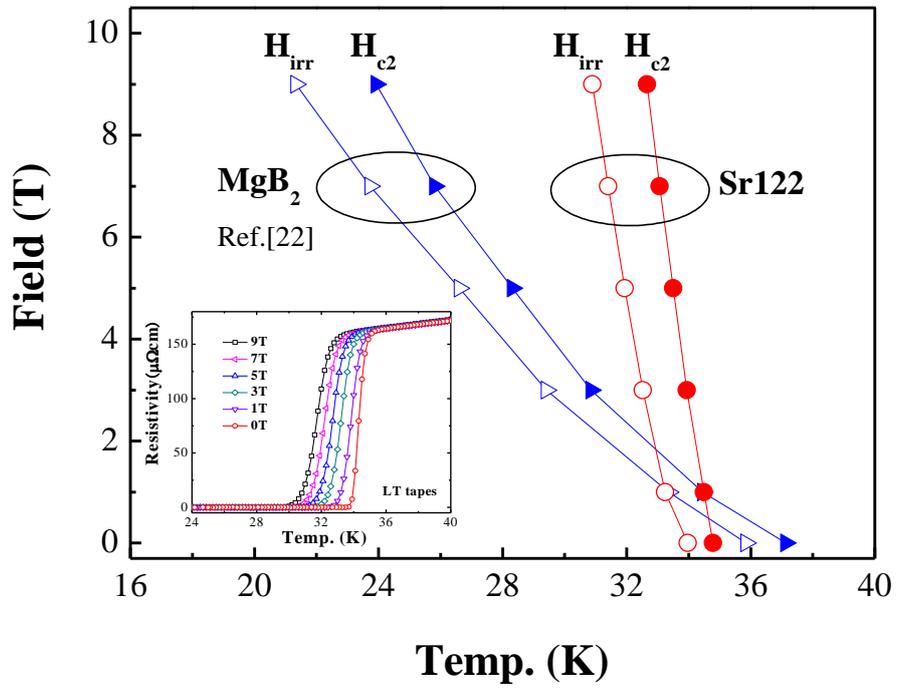

Figure 5    Ma et al.